\hsize=15.5 truecm
\vsize=22.5 truecm
\leftskip=1 truecm
\topskip=1 truecm
\splittopskip=1 truecm
\parskip=0 pt plus 1 pt
\baselineskip=18.9 pt
\def\re{$\rm I \kern-0.07cm R$}
\def\a{\alpha}
\def\b{\beta}
\def\g{\gamma}

\def\ve{\varepsilon}

%\hfill {\it ESI Preprint Series No. 168/94}
\bigskip
\bigskip
\bigskip
\centerline{\bf ON EINSTEIN'S EQUATIONS FOR SPACETIMES ADMITTING}
\centerline{\bf A NON-NULL KILLING FIELD}
\bigskip
\bigskip
\centerline{{Istv\'an R\'acz}\footnote\dag
{\sevenrm{Email: istvan@rmkthe.rmki.kfki.hu}}}
\medskip
\centerline{MTA KFKI Research Institute for Particle and Nuclear
Physics}
\centerline{H-1525 Budapest 114, P.O.B. 49, Hungary }
\bigskip
\bigskip
\bigskip
\parindent 0pt
 
{\bf Abstract:} We consider the 3-dimensional formulation of Einstein's
theory for spacetimes possessing a non-null Killing field $\xi^a$.  It is
known that for the vacuum case some of the basic field equations are
deducible from the others.  It will be shown here how this result can be
generalized for the case of essentially arbitrary matter fields.  The
systematic study of the structure of the fundamental field equations is
carried out.  In particular, the existence of geometrically preferred
reference systems is shown. Using local coordinates of this type
two approaches are
presented resulting resolvent systems of partial differential equations for the
basic field variables.  Finally, the above results are applied
for perfect fluid spacetimes describing possible equilibrium configurations
of relativistic dissipative fluids.
 
\par
 
\parindent 20pt
\bigskip
\bigskip

\parindent 0 pt\bigskip
{\bf I. Introduction}
\parindent 20 pt\medskip
 
If one is given a complicated system of nonlinear partial differential
equations to solve -- as it happens frequently, for instance, in
Einstein's gravitational theory -- it is hard to see whether there
exists any relationship between the equations or not.  Sometimes the
realization of certain type of connection might induce the introduction
of an entirely new technique in solving the selected problem.  This was
the case, for example, when Cosgrove [1] gave a new formulation of field
equations for stationary axisymmetric vacuum gravitational fields, or
when, by a generalization of Cosgrove's approach, Fackerell and Kerr [2]
derived a resolvent system of differential equations for the vacuum
field equation of Einstein's theory for spacetimes with a single
non-null Killing vector field.
 
In the first part of this paper we are going to show that the fundamental
results the introduction of the new approach was based on in Refs. [1,2]
can be generalized for spacetimes possessing a non-null Killing field with
essentially arbitrary matter fields. Subsequently, the properties of the
basic field equations are studied in the situation where the gradient of
the norm of the Killing field and the twist of the Killing field are
linearly independent.  It is shown, for instance, that there exists a
geometrically preferred vector field on the space of Killing orbits so
that the basic field equations possess -- in local coordinate systems
adopted to this vector field -- very simple form.  In particular, a number
of the relevant field variables and/or their partial derivatives with
respect to the coordinate associated with the preferred vector field are
found to be identically zero.  By the application of the associated
simplifications, two different approaches in deriving resolvent systems of
partial differential equations for the basic field variables are presented. The
first is a general approach while the second one is a generalization,
for particular matter fields, of
the techniques applied for the study of stationary axisymmetric vacuum
fields by Cosgrove [1]. In the last part of this paper the application
of both of these techniques for the case of perfect fluids
possessing 4-velocity parallel to a timelike Killing
field will be
presented.
 
\parindent 0 pt\bigskip
{\bf II. The field equations}
\parindent 20 pt\medskip
 
In this section, first, we shall recall some of the notions and
techniques of the formalism of general relativity developed for
spacetimes possessing a non-null Killing vector field.  Then, it will be
shown that some of the field equations involved are always deducible
from the others.  Finally, as a direct application of this result the
basic field equations will be reformulated -- displaying the simplest
form of the relevant equations -- corresponding to the possible
subcases.
 
Consider a  smooth spacetime, $(M,g_{ab})$, with a non-null Killing vector field, 
$\xi^a$. It is well known that for such a spacetime the formulation of 
the Einstein's theory can be simplified considerably by making use of a
3-dimensional formalism [3,4].  In particular, this is done as follows: 
Let $\cal S$ denote the space of Killing orbits of $\xi^a$.  It is assumed
here that $\cal S$ can be given the structure of a 3-dimensional
differentiable manifold so that the projection map,
$\phi:M\rightarrow{\cal S}$, from $M$ onto $\cal S$ is a smooth mapping
[3].  This condition always holds locally, and for the case of a
timelike Killing field in a chronological spacetime is shown to be
satisfied globally [5].  Consider, now, the following three fields on $M$: the
norm of the Killing field $$v= \xi^a\xi_a,\eqno(2.1)$$ the twist of the
Killing field $$\omega_a=\epsilon_{abcd}\
\xi^b\nabla^c\xi^d,\eqno(2.2)$$ and the symmetric tensor field
$$h_{ab}=g_{ab}-v^{-1}\xi_a\xi_b .\eqno(2.3)$$ The images of these 
fields by the differential, $\phi_*$, of $\phi$ give
rise to tensor fields on the 3-space $\cal S$.  For instance,
$\phi_*h_{ab}$ is the natural induced 
metric on $\cal S$ which is Lorentzian or Riemannian according to 
that the Killing field,
$\xi^a$, is spacelike or timelike.  (Hereafter we restrict our
considerations to the 3-space $\cal S$ so it should not cause a big
confuse that the same notation will be used for the tensor fields living
on $\cal S$ and for their natural `pull backs' onto $M$.)
 
Then the basic field equations are
[4,5] $$R_{ab}^{ ^{(3)}}={1\over 2}v^{-1}D_aD_bv-{1\over
4}v^{-2}(D_av)(D_bv)+ {1\over
2}v^{-2}\{\omega_a\omega_b-h_{ab}(\omega_m\omega^m)\}+ h_a^mh_b^n R_{mn}^{
^{(4)}}, \eqno(2.4)$$ $$D_{[a}\omega_{b]}=-\epsilon_{abmn} \xi^m h^n_p R^{
^{(4)}p}_q\xi^q, \eqno(2.5) $$ $$D^aD_av={1\over 2}v^{-1}(D_mv)(D^mv)-
v^{-1}\omega_m\omega^m- 2 R_{mn}^{ ^{(4)}}\xi^m\xi^n, \eqno(2.6) $$
$$D^a\omega_a={3\over 2}v^{-1}\omega_mD^mv, \eqno(2.7) $$ where
$R_{ab}^{^{(3)}}$ and $D_a$ denote the Ricci tensor and the covariant
derivative operator associated with $h_{ab}$, while, $R_{ab}^{^{(4)}}$ is
supposed to be given in terms of the energy-momentum tensor, $T_{ab}$, of
the matter fields by virtue of Enstein's equations $$R_{ab}^{^{(4)}}=8\pi
(T_{ab}-{1\over 2} g_{ab} T).\eqno(2.8)$$ Equations (2.4) - (2.7) relate
the various type of projections of the 4-Ricci tensor to tensor fields and
their covariant derivatives living on the 3-space $\cal S$. It is
important that the entire geometrical content of Einstein's
theory for a spacetime, $(M,g_{ab})$, with a non-null Killing vector
field, $\xi^a$, can be uniquely represented by a 3-dimensional metric
space, $({\cal S},h_{ab})$, along with the fields $v$ and $\omega_a$
satisfying the above set of field equations. Even more important that, to
any 3-dimensional formulation, $\{({\cal S},h_{ab});v,\omega_a\}$, of this
type -- up to gauge transformations -- there exists a unique 4-dimensional
spacetime, $(M,g_{ab})$, with a Killing field, $\xi^a$, so that the
projection map $\phi:M\rightarrow{\cal S}$ reproduces the 3-dimensional
formulation we started with.  In fact, (2.7) is just the integrability
condition ensuring that the 4-geometry can be recovered from the
3-dimensional formulation [3,4].
 
Note that no restrictions have been raised concerning the matter fields.
In fact, what we really need is that the matter
fields be represented by tensor fields, ${{\Psi_{_{(i)}}}^{a_1 ...
a_k}}_{b_1 ... b_l}$ ($i\in {\cal I}$), on $M$, and, a diffeomorphism
invariant action be associated with them so that the energy-momentum
tensor, $T_{ab}$, and the Euler-Lagrange equations can be expressed in
terms of appropriate variations of this action.
 
It is important to note that the invariance of $T_{ab}$ under the
action of the isometry group associated with $\xi^a$ do not imply
that the fields
${{\Psi_{_{(i)}}}^{a_1...a_k}}_{b_1...b_l}$
are invariant. There are, for instance, exact solutions of the stationary
vacuum Einstein-Maxwell field equations so that the electromagnetic fields
are non-stationary [6]. On
the other hand, whenever
${\cal L}_\xi{{\Psi_{_{(i)}}}^{a_1...a_k}}_{b_1...b_l}=0$ for each value
of $i$ one might consider the unique decomposition of the fields
${{\Psi_{_{(i)}}}^{a_1...a_k}}_{b_1...b_l}$
into tensor fields which possess definite `tangential' or `perpendicular'
character with regard to their free indices. These fields can be built up
from tensorial products of $\xi^a$, $\xi_a$ and the pull backs of tensor fields
${{\psi_{_{(j)}}}^{a_1...a_m}}_{b_1...b_n}$
living on $\cal S$.
 
It is well known that equations (2.4) - (2.7) can be simplified by the
introduction of the conformal metric $\hat h_{ab}$ defined as $$\hat
h_{ab}=\ve v h_{ab},\eqno(2.9)$$ where $\ve$ takes the value $+1$ (resp.
$-1$) for spacelike (resp. timelike) Killing fields.  Then (2.4) - (2.7)
take the form $$\hat R_{ab}={1\over 2}v^{-2}\{(\hat D_av)(\hat D_bv)+
\omega_a\omega_b\}+\{ h_a^m h_b^n+\ve v^{-2}\hat h_{ab}\xi^m\xi^n\}
R_{mn}^{ ^{(4)}}, \eqno(2.10) $$ $$\hat D_{[a}\omega_{b]}=
-\epsilon_{abmn} \xi^m R^{ ^{(4)}n}_p\xi^p, \eqno(2.11) $$ $$\hat
D^a\hat D_av=v^{-1}\{(\hat D_mv)(\hat D^mv)- \omega_m\omega^m\}- 2\ve
v^{-1} R_{mn}^{ ^{(4)}}\xi^m\xi^n, \eqno(2.12) $$ $$\hat D^a\omega_a=
2v^{-1}\omega_m\hat D^mv, \eqno(2.13) $$ where $\hat D_a$ and $\hat
R_{ab}$ are the covariant derivative operator and the Ricci tensor
associated with $\hat h_{ab}$.
 
\medskip
 
 Although we are considering the set of basic field equations for
spacetimes with a Killing vector field -- in which case some
simplification arise compared to the general case -- the whole set of
field equations is still rather complicated.  For instance, equations
(2.10) - (2.13) give rise -- in local coordinates -- to a system of
coupled non-linear second order partial differential equations for the
function $v$ and the components of the tensor fields $\omega_a, {\hat
h}_{ab}$, ${{\Psi_{_{(i)}}}^{a_1 ... a_k}}_{b_1 ... b_l}$.  In fact, the
situation is, in general, even worse because, in addition to (2.10) -
(2.13), we have to solve simultaneously the Euler - Lagrange equations
which govern the evolution of matter fields in the spacetime.  Note that
these equations of motion, in general, couple to the above set of field
equations increasing thereby the complexity of the whole problem.
Therefore it is important to know what are the exact relationships
between these equations.
 
\medskip
 
Now, we are going to show that in the formulation of Einstein's theory
for spacetimes possessing a non-null Killing vector field the same type
of simplification arises as for the case of stationary axisymmetric
vacuum case realized by Cosgrove [1].  In particular, it can be shown
that equations (2.10) and (2.11) are actually far more fundamental than
(2.12) and (2.13).  More precisely, by using (2.10) and (2.11) one can
derive the following algebraic relationship $$\eqalign{ (\hat
D_bv)\Bigl[\hat D^a\hat D_av-&v^{-1}\{(\hat D_mv)(\hat D^mv)-
\omega_m\omega^m\}+ 2\ve v^{-1} R_{mn}^{ ^{(4)}}\xi^m\xi^n\Bigr]+
\omega_b\Bigl[\hat D^a\omega_a- 2v^{-1}\omega_m\hat D^mv\Bigr]\cr&+ \ve
v^{-1}{h_b}^m\bigl[\nabla^n {R}^{ ^{(4)}}_{mn}-{1\over 2}\nabla_m
R^{^{(4)}} \bigr]=0.\cr} \eqno(2.14)$$

Up to this point we have considered only the set of basic field equation 
(2.10) - (2.13) which are equivalent to Einstein's equations and an 
integrability condition. Remember that the 4-Ricci tensor, $R_{ab}^{ ^{(4)}}$ 
was assumed to be given in terms of the energy-momentum tensor, $T_{ab}$.
Thereby, the last term of the left hand side of (2.14), which is, in fact,
$\varepsilon v^{-1} {h_b}^m\nabla^n T_{m n}$, cannot be put zero simply by 
referring to the 4-dimensional twice contracted Bianchi identity. It is known, 
however, that this term is identically zero whenever either the complete set 
of Einstein's equations or the Euler-Lagrange equations for the matter fields 
are satisfied. Since our aim is to derive a relationship between some of the 
relevant Einstein's equations to get rid of this term later it will be assumed 
that the equations of motion are satisfied by matter fields.
[The author would like to say thank you to the unknown referee who 
pointed out the need for the clarification why the 4-dimensional Bianchi 
identity cannot be applied to set the third term of (2.14) to zero immediately.]
 
\baselineskip=18.9 pt

The way one could get the relation (2.14) is the following: Substitute the
right hand side of (2.10) for $\hat R_{ab}$ into the following
expression $${\hat D}^a{\hat R}_{ab} -{1\over 2} {\hat D}_b {\hat
R}.\eqno(2.15)$$ Then by using (2.11) a straightforward calculation
yields that $$\eqalign{{\hat D}^a{\hat R}_{ab}-{1\over 2} {\hat D}_b
{\hat R}=&{1\over 2} v^{-2}\Biggl\{(\hat D_bv)\Bigl[\hat D^a\hat
D_av-v^{-1}\{(\hat D_mv)(\hat D^mv)- \omega_m\omega^m\}+ 2\ve v^{-1}
R_{mn}^{ ^{(4)}}\xi^m\xi^n\Bigr]\cr& \phantom{ {1\over 2}v^{-2}\Biggl\{
} +\omega_b\Bigl[\hat D^a\omega_a- 2v^{-1}\omega_m\hat
D^mv\Bigr]\Biggr\}\cr&-\ve v^{-3} (\hat D_bv) (R_{mn}^{
^{(4)}}\xi^m\xi^n)+ v^{-2}\omega^a\hat D_{[a}\omega_{b]}+\Bigl[ {\hat
D}^a{\rho}_{ab}-{1\over 2} {\hat D}_b({\hat h}^{mn} \rho_{mn})\Bigr],
\cr}\eqno(2.16)$$ where $\rho_{ab}=\{ h_a^m h_b^n+\ve v^{-2}\hat
h_{ab}\xi^m\xi^n\} R_{mn}^{ ^{(4)}}.$ Since $\xi^a$ is a Killing field
on $M$ we get by (2.13) $$v^{-2}\omega^a\hat D_{[a}\omega_{b]}= 2\ve
v^{-2} {h_b}^m(\nabla_n\xi_m) R^{^{(4)}n}_p\xi^p+\ve v^{-3} (\hat D_bv)
(R_{mn}^{ ^{(4)}}\xi^m\xi^n).\eqno(2.17)$$ We also have, for instance,
${\cal L}_\xi R^{^{(4)}}_{ab}=0$.  Moreover, it can be shown by using
the relationship between the covariant derivative operators $D_a$ and
$\hat D_a$ -- with a tedious but straightforward calculation -- that
$${\hat D}^a{\rho}_{ab}-{1\over 2} {\hat D}_b({\hat h}^{mn} \rho_{mn})=
\ve v^{-1}{h_b}^m\bigl[\nabla^n {R}^{ ^{(4)}}_{mn}-{1\over 2}\nabla_m
R^{^{(4)}} \bigr] - 2\ve v^{-2}{h_b}^m(\nabla_n\xi_m) R^{
^{(4)}n}_p\xi^p.\eqno(2.18)$$
 
Now using (2.16),(2.17) and (2.18) we obtain $$\eqalign{{\hat D}^a{\hat
R}_{ab}-{1\over 2} {\hat D}_b {\hat R}= &{1\over 2} v^{-2}\Biggl\{(\hat
D_bv)\Bigl[\hat D^a\hat D_av-v^{-1}\{(\hat D_mv)(\hat D^mv)-
\omega_m\omega^m\}+ 2\ve v^{-1} R_{mn}^{
^{(4)}}\xi^m\xi^n\Bigr]\cr&+\omega_b\Bigl[\hat D^a\omega_a-
2v^{-1}\omega_m\hat D^mv\Bigr]+ \ve v^{-1}{h_b}^m\bigl[\nabla^n {R}^{
^{(4)}}_{mn}-{1\over 2}\nabla_m R^{^{(4)}} \bigr] \Biggr\}.
\cr}\eqno(2.19)$$ Since the tensor field $\hat R_{ab}$ is just the Ricci
tensor associated with the three metric, $\hat h_{ab}$, -- in virtue of the
twice contracted Bianchi identity -- we have that the left hand side of the
previous equation is identically zero.  This proves then that (2.14)
holds identically.
 
\medskip In the remaining part of this section we are going to study the
consequences of the algebraic relation (2.14).  We shall use the
assumption that the Euler-Lagrange equations are satisfied for matter
fields which implies in the case when they are derived from a diffeomorphism
invariant action that $$\nabla^a T_{ab}=0 \eqno(2.20)$$ so the
third term of (2.14) is zero.  Therefore, we have that the relevant form
of (2.14) says then that the above particular linear combination of the
form fields, $\hat D_av$ and $\omega_a$, must vanish identically.
Correspondingly, there are two subcases which have to be treated
separately, namely, $\hat D_a v$ and $\omega_a$ might be either linearly
independent or not.
 
 \medskip Whenever the two form fields $\hat D_av$ and $\omega_a$ are
linearly independent only the trivial combinations of them can vanish
identically.  In this case (2.12) and (2.13) can be deduced from (2.10),
(2.11) and the Euler-Lagrange equations.  It is then sufficient to solve
(2.10) and (2.11) along with the relevant equations of motion for matter
fields since any solution of these equations will automatically satisfy
(2.12) and (2.13) as well.
 
\medskip
 Suppose now that the two form fields, $\hat D_av$ and
$\omega_a$, are linearly dependent.
This might happen whenever one of them vanishes throughout or
there exists a function, $f$, such that
$$\omega_a=f\cdot(\hat D_a v).\eqno(2.21)$$
 
\medskip
 
$\alpha$; Consider first the case of vanishing $\hat D_a v$, i.e., we
suppose that $v$ is constant throughout.  Since we can introduce then a
new Killing field instead of $\xi^a$ by rescaling $\xi^a$ with an
arbitrarily chosen constant factor we may assume here, without loss of
generality, that $v=\ve$.  Furthermore, for this case (2.14) implies that
the relevant form of (2.13) is a consequence of (2.10), (2.11) and (2.20).
Hence, the whole content of the basic field equations reduce to $$\hat
R_{ab}={1\over 2} \omega_a\omega_b+\{ h_a^m h_b^n+\ve \hat
h_{ab}\xi^m\xi^n\} R_{mn}^{ ^{(4)}}, \eqno(2.\alpha.1) $$ $$
\omega_m\omega^m= - 2\ve R_{mn}^{ ^{(4)}}\xi^m\xi^n, \eqno(2.\alpha.2)$$
$$\hat D_{[a}\omega_{b]}= -\epsilon_{abmn} \xi^m R^{ ^{(4)}n}_p\xi^p.
\eqno(2.\alpha.3) $$
 
 \medskip $\beta$; Suppose now that $\omega_a=0$, i.e., $\xi^a$ is
hypersurface orthogonal.  Then  (2.11) and (2.13) are  expected
to hold, furthermore, the relevant  form of    (2.12) is  simply a
consequence of     (2.14).   Hence, the basic  equations for  the
case  under  consideration simplify to [4] $$\hat  R_{ab}={1\over
2}
v^{-2}(\hat  D_av)(\hat  D_bv)+\{  h_a^m  h_b^n+\ve  v^{-2}   \hat
h_{ab}\xi^m\xi^n\} R_{mn}^{ ^{(4)}}. \eqno(2.\beta) $$

\medskip
 
$\gamma$; Finally, suppose that neither $\hat D_av$ nor $\omega_a$
vanishes, and, there exists a function, $f$, such that (2.21) holds.  Then
the elimination of $\omega_a$ from (2.10) - (2.13) yields by using the
above relationship $$\hat R_{ab}={1\over 2}v^{-2}(1+f^2)(\hat D_av)(\hat
D_bv)+\{ h_a^m h_b^n+\ve v^{-2}\hat h_{ab}\xi^m\xi^n\} R_{mn}^{ ^{(4)}},
\eqno(2.\gamma.1) $$ $$\hat D^a\hat D_av=v^{-1}(1-f^2)(\hat D_mv)(\hat
D^mv) - 2\ve v^{-1} R_{mn}^{ ^{(4)}}\xi^m\xi^n, \eqno(2.\gamma.2) $$
$$(\hat D_a f)(\hat D^a v)= v^{-1}[(1+f^2)(\hat D_m v)(\hat D^mv)+2\ve
R_{mn}^{^{(4)}}\xi^m\xi^n], \eqno(2.\gamma.3)$$ $$(\hat D_{[a}f)(\hat
D_{b]}v)= -\epsilon_{abmn} \xi^m R^{ ^{(4)}n}_p\xi^p. \eqno(2.\gamma.4) $$
It can easily be checked that the relevant form of (2.14) implies that
(2.$\gamma$.2) is deducible from the Euler-Lagrange equations and
(2.$\gamma$.1), (2.$\gamma$.3) and $(2.\gamma$.4).  Hence, for this last
case, equations (2.21), (2.$\gamma$.1),(2.$\gamma$.3) and (2.$\gamma$.4)
display the entire content of the basic field equations.

\bigskip
\parindent 0 pt\bigskip
{\bf III. Geometrically preferred local coordinates }
\parindent 20 pt\medskip
 
In the remaining part of this paper we shall restrict our consideration to
the case of independent form fields, i.e., we suppose that $(\hat
D_{[a}v)\omega_{b]}\not=0$ on a subset $\tilde {\cal S}$ of ${\cal S}$.  (The
other possibility, when $\hat D_av$ and $\omega_b$ are linearly dependent,
will be examined elsewhere.) According to the results of the previous
section to get a solution of the basic field equations, (2.10) - (2.13),
it is sufficient to solve (2.10) and (2.11) along with the relevant set of
Euler-Lagrange equations.  In this section we are going to examine the
properties of the fundamental equations (2.10) and (2.11).  In particular,
it will be shown that there exist geometrically preferred local coordinate
systems in which these equations possess very simple form.
 
We shall use the following shortened form of (2.10) and (2.11) $$\hat
R_{ab}={1\over 2}v^{-2}\{(\hat D_av)(\hat D_bv)+
\omega_a\omega_b\}+\rho_{ab} , \eqno(3.1) $$ $$\hat D_{[a}\omega_{b]}=
\sigma_{ab} , \eqno(3.2) $$ where $$\rho_{ab}=\{ h_a^m h_b^n+\ve
v^{-2}\hat h_{ab}\xi^m\xi^n\} R_{mn}^{ ^{(4)}},\eqno(3.3)$$ and
$$\sigma_{ab}=
-\epsilon_{abmn} \xi^m R^{ ^{(4)}n}_p\xi^p. \eqno(3.4)$$
Note that $\rho_{ab}$ is a
symmetric while $\sigma_{ab}$ an antisymmetric tensor field on $\cal
S$, both depending on the fields $v,\hat h_{ab},$ ${{\Psi_{_{(i)}}}^{a_1
... a_k}}_{b_1 ... b_l}$.
 
Since  $(\hat D_{[a}v)\omega_{b]}\not=0$ on $\tilde {\cal S}$
there exists a nowhere vanishing vector field, $k^a$, there defined
as $$k^a=\hat \epsilon^{abc}(\hat D_{b}v)\omega_{c},\eqno(3.5)$$
where $\hat \epsilon_{abc}$ denotes the 3-dimensional volume
element associated with $\hat h_{ab}$, i.e.,  $\hat \epsilon_{abc}=
\epsilon_{abcd}\xi^d$.
 
Then the following hold
$${\cal L}_k v=0,\eqno(3.6)$$
$$k^a\omega_a =0,\eqno(3.7)$$
$$k^a(\hat R_{ab}-\rho_{ab})=0,\eqno(3.8)$$
and
$${\cal L}_k (\hat
R_{ab}-\rho_{ab})=v^{-2}k^e\{\sigma_{ea}\omega_b+\sigma_{eb}\omega_a\}.
\eqno(3.9)$$
 
Equations (3.6) - (3.8) are direct consequences of the
definition of $k^a$. For   (3.9) note that
$${\cal L}_k(\hat R_{ab}-\rho_{ab})=k^e\hat D_e(\hat R_{ab}-\rho_{ab})
+(\hat R_{eb}-\rho_{eb}) \hat D_a k^e+
(\hat R_{ae}-\rho_{ae}) \hat D_b k^e.\eqno(3.10)$$
However, according to   (3.8) we have $(\hat
R_{eb}-\rho_{eb})\hat D_a
k^e= - k^e \hat D_a (\hat R_{eb}-\rho_{eb})$, and so
$${\cal L}_k(\hat R_{ab}-\rho_{ab})=k^e\bigl\{\hat D_e(\hat
R_{ab}-\rho_{ab})-\hat D_a (\hat R_{eb}-\rho_{eb}) -
\hat D_b (\hat R_{ae}-\rho_{ae}) \bigr\}.\eqno(3.11)$$
Now, using   (3,1),(3.6) and (3.7) we get
$${\cal L}_k(\hat R_{ab}-\rho_{ab})=v^{-2} k^e\bigl\{
(\hat D_{[e}\hat D_{a]}v)(\hat D_bv)+
(\hat D_av)(\hat D_{[e}\hat D_{b]}v)+
\hat D_{[e}\omega_{a]}\omega_b+\omega_a\hat
D_{[e}\omega_{b]}\bigr\},\eqno(3.12)$$
which imply, along with   (3.2) and the fact that $\hat D_a$ is
torsion free, that   (3.9) holds.

\parindent 20 pt
\bigskip
 
Note that whenever $k^e \sigma_{ea}$ is vanishing on $\tilde {\cal S}$, we have
$${\cal L}_k(\hat R_{ab}-\rho_{ab})=0, \eqno(3.13)$$ i.e., $k^a$ is a
collineation vector field of $\hat R_{ab}-\rho_{ab}$.  According to the
definition of $\sigma_{ab}$ the contraction $k^e\sigma_{ea}$ is
identically zero whenever there exist functions $\alpha,\beta$ such that
$R^{^{(4)}n}_p\xi^p=\alpha \xi^n + \beta k^n.$ For the vacuum case,
$\rho_{ab}=\sigma_{ab}=0$ (or $\a=\b=0$). Then (3.9) reduces to the
well-known result that $k^a$ is a Ricci collineation vector [7].

It is straightforward to check that 
(3.6) -- (3.9) are satisfied not merely for $k^a$ but for any vector field 
possessing the form $f k^a$, where $f$ is an arbitrary function on 
$\tilde {\cal S}$. Hereafter, 
the vector fields, $\hat k^a=f k^a$, which are defined 
with the use of a non-vanishing function, $f$, on 
$\tilde {\cal S}$ will be
referred as being geometrically preferred.

\medskip
 
Just like for the vacuum case (see Ref. [2]) one can introduce geometrically
preferred local coordinate systems. Denote by $\hat k^a$  any of the 
geometrically preferred vector fields
and consider local
coordinates, $(x^1,x^2,x^3)$, adopted to $\hat k^a$, i.e.,
$$\hat k^a=\Bigl({\partial\over\partial x^3}\Bigl)^a, \ \ \ {\rm  or}\ \ \
\hat k^\alpha= {\delta_3}^\alpha.\eqno(3.14)$$ This type of coordinates
can always be introduced (at least locally) on $\tilde
{\cal S}$.
 
In such a local coordinate system, $(x^1,x^2,x^3)$, equations (3.6) -
(3.9) take the form $${\partial v\over\partial x^3}=0,\eqno(3.15)$$
$$\omega_3 =0,\eqno(3.16)$$ $$\hat
R_{3\beta}-\rho_{3\beta}=0,\eqno(3.17)$$ and $${\partial\over \partial
x^3} (\hat R_{\alpha\beta}-\rho_{\alpha\beta})
=v^{-2}\{\sigma_{3\alpha}\omega_ \beta+\sigma_ {3\beta}\omega_\alpha\}
\eqno(3.18)$$ where $\beta$ takes the values $1,2,3$.  Note that whenever
one of the functions, $\rho_{3\b}$ $(\b=1,2,3)$, does not vanish
identically (3.17) gives algebraical relationship(s) between the variables
$v,\hat h_{ab},$ ${{\Psi_{_{(i)}}}^{a_1 ... a_k}}_{b_1 ... b_k}$ and
the derivatives of $\hat h_{ab}$ and ${{\Psi_{_{(i)}}}^{a_1 ...
a_k}}_{b_1 ... b_k}$.  Then we get that in such an adopted local
coordinate system, $(x^1,x^2,x^3)$, (3.1) is equivalent to (3.17) and
$$\hat R_{AB}={1\over 2}v^{-2}\{(\partial_Av)(\partial_Bv)+
\omega_A\omega_B\}+\rho_{AB}, \eqno(3.19)$$ where $\partial_Av$ denotes
the partial derivative of $v$ with respect to the variable $x^A$, and the
capital Latin indices take the values $1,2$.
 
It can be easily checked that in such a coordinate
system (3.2) takes the form $$\eqalign{\partial_1
\omega_2-\partial_2\omega_1=& \sigma_{12}\cr \partial_3
\omega_A=&\sigma_{3A}.\cr}\eqno(3.20)$$ Observe that, whenever $\sigma_{ab}$
vanishes identically these equations imply that (at least locally) there
exists a function, $\omega=\omega(x^1,x^2)$, so that
$$\omega_A=\partial_A
\omega.\eqno(3.21)$$
 
Using these simplifications, in the next two sections two different
methods in establishing resolvent systems of partial differential
equations for the
basic field variables will be presented.
 
\parindent 0 pt\bigskip
{\bf IV. General method}
\parindent 20 pt\medskip
 
This section is devoted to the introduction of a general approach to get a
resolvent system of differential equations for the basic field
variables. This approach is based on the following observation:
It seems to be a general feature of the present formulation of Einstein's theory
that (3.17) can be solved for the function $v$
in many cases.
Hereafter, we shall assume that the fields
${{\Psi_{_{(i)}}}^{a_1...a_k}}_{b_1...b_l}$
are invariant under the action of the isometry group associated with $\xi^a$,
thereby, we can use the fields
${{\psi_{_{(j)}}}^{a_1...a_m}}_{b_1...b_n}$ to represent the
matter content, instead of them.
Combining these two facts, hereafter we shall assume that the norm of
the Killing
field, $v$, can be given in terms of quantities derived from the
induced 3-geometry, $\hat h_{ab}$,
and, possibly, from the tensor fields
${{{\psi_{_{(i)}}}^{a_1...a_m}}}_{b_1...b_n}$, representing the matter
fields.
Correspondingly, we shall assume that  there exists a function
$$v=v({\hat h}_{\a\b},\ {\partial_\delta}{\hat h}_{\a\b},
\ \partial_\delta{\partial_\rho}
{\hat h}_{\a\b};\
{{{\psi_{_{(j)}}}^{\a_1...\a_m}}}_{\b_1...\b_n},\
\partial_\delta{{{\psi_{_{(j)}}}^{\a_1...\a_m}}}_{\b_1...\b_n}, \
\partial_\delta\partial_\rho{{{\psi_{_{(j)}}}^{\a_1...\a_m}}}_{\b_1...\b_n})
,\eqno(4.1)$$
where the presence of second order partial derivatives of the fields
${{{\psi_{_{(j)}}}^{\a_1...\a_m}}}_{\b_1...\b_n}$ indicates that
the matter Lagrangian is supposed to contain at most second order partial
derivatives of these fields
and the Greek indices refer to components of tensor fields in
geometrically preferred
adopted local coordinates.
 
To start off note that (3.19) can be recast into the form
$$H_{AB}=v^{-2}\{(\partial_Av)(\partial_Bv)+ \omega_A\omega_B\},
\eqno(4.2)$$ where $$H_{AB}=2 (\hat R_{AB}-\rho_{AB}).\eqno(4.3)$$
It is important to emphasize that at each (explicit or implicit)
appearance of the function
$v$ in (4.2) the substitution of
the right hand side of (4.1) is understood. Since we are dealing with the
case of linearly independent form fields, i.e.,
$(\hat D_{[a}v)\omega_{b]}\not=0$ on $\tilde {\cal S}$,
(4.2) can be shown to be
equivalent to
the following set of equations
$$\omega_A=\epsilon{H_{A2}(\partial_1v)-H_{1A}(\partial_2v) \over
[det(H_{AB})]^{1\over 2}},\eqno(4.4)$$
and $$H_{11}(\partial_2v)^2-2 H_{12}
(\partial_1 v)(\partial_2 v) + H_{22}(\partial_1 v)^2-v^2
det(H_{AB})=0,\eqno(4.5)$$ where
the sign ambiguity of $\omega_A$
is indicated by the factor $\epsilon$ (i.e., $\epsilon=\pm 1$) in (4.4).
 
Using the definition, (3.3), of $\rho_{ab}$ and (3.17) it can be
checked easily that $v$ depends on at most second order derivatives of the
metric
functions,
$\hat h_{\a\b}$, since only the terms $\hat R_{\a 3}$ enter (3.17).
Therefore, with the assumption that at most second
order covariant derivatives of the fields
${{{\psi_{_{(j)}}}^{a_1...a_m}}}_{b_1...b_n}$ are involved in the matter
Lagrangian, we can conclude that
(4.5) is at most
a third order partial differential equation for the fields $\hat
h_{\alpha\beta}$ and
${{{\psi_{_{(j)}}}^{\alpha_1...\alpha_m}}}_{\beta_1...\beta_n}$. Three
additional partial
differential equations restricting  these fields have to be taken
into consideration. These are derived by
substituting the right hand side of (4.4) for $\omega_A$ into (3.20) and
can be given as follows:
$$\eqalign{&
(\partial_1 H_{22})(\partial_1 v)+H_{22}(\partial_1\partial_1 v)-
(\partial_1 H_{12})(\partial_2 v)-2 H_{12}(\partial_2\partial_1 v)-
(\partial_2 H_{12})(\partial_1 v)\cr&+
(\partial_2 H_{11})(\partial_2 v)+H_{11}(\partial_2\partial_2 v)+
\partial_1\Bigl(ln[det(H_{AB})]^{-{1\over 2}}\Bigr)
\Bigl(H_{22}(\partial_1 v)-H_{12}(\partial_2 v)\Bigr)\cr&\phantom{
+(\partial_2 H_{11})(\partial_2 v)}-
\partial_2\Bigl(ln[det(H_{AB})]^{-{1\over 2}}\Bigr)\Bigl(H_{12}(\partial_1v)
-H_{11}
(\partial_2v)\Bigr)=\epsilon [det(H_{AB})]^{{1\over 2}} \sigma_{12},\cr}
\eqno(4.6)$$
$$\eqalign{(\partial_3 H_{12})(\partial_1 v)-
(\partial_3 H_{11})(\partial_2 v)+
\partial_3\Bigl(ln[det(H_{AB})]^{-{1\over 2}}\Bigr)
\Bigl(H_{12}(\partial_1 v)&-H_{11}(\partial_2 v)\Bigr)\cr&
=\epsilon [det(H_{AB})]^{{1\over 2}} \sigma_{31},\cr}
\eqno(4.7)$$
$$\eqalign{(\partial_3 H_{22})(\partial_1 v)-
(\partial_3 H_{21})(\partial_2 v)+
\partial_3\Bigl(ln[det(H_{AB})]^{-{1\over 2}}\Bigr)
\Bigl(H_{22}(\partial_1 v)&-H_{12}(\partial_2 v)\Bigr)\cr&
=\epsilon [det(H_{AB})]^{{1\over 2}} \sigma_{32}.\cr}
\eqno(4.8)$$
 
The first equation, (4.6), is a fourth order while the last two ones
are third order non-linear partial differential equations.
These equations along with (3.17), (4.5) and
the relevant set of Euler-Lagrange equations give rise to a resolvent
system of
field equations for the variables $\hat h_{\alpha\beta}$ and
${{{\psi_{_{(j)}}}^{\alpha_1...\alpha_m}}}_{\beta_1...\beta_n}$.
Once one could get a solution of these field equations one can determine
$v$ via
(4.1), moreover, $\omega_A$ can be given
in virtue of
(4.4).
 
Clearly, the applicability of this approach strongly depends on the detailed
functional form of $v$ which was implicitly used throughout this section.
For instance,
the explicit form of the basic field equations, (3.17), (4.5)-(4,8),
for the variables $\hat h_{\a\b}$
and ${{{\psi_{_{(j)}}}^{\alpha_1...\alpha_m}}}_{\beta_1...\beta_n}$ can
be examined only for particular matter fields separately. In section 7 we are
going to give the functional form of $v$ for perfect
fluid matter sources possessing 4-velocity parallel to a timelike
Killing field and for particular equations of state.
 
\parindent 0 pt\bigskip
{\bf V. Generalization of Cosgrove's method}
\parindent 20 pt\medskip
 
 In this section we generalize the techniques developed originally
for   stationary   axisymmetric   vacuum   fields   for spacetimes
possessing  a  singe  non-null  Killing  field  with matter fields
satisfying   the   additional   conditions   given   below.   More
precisely,  a  slightly  modified  version of Cosgrove's approach
will  be  established  so as to derive from the basic set of field
equations a resolvent  system of differential equations  for
the basic variables.
 
The two conditions are the following:
\medskip
\parindent 0 pt
 
\leftskip=3.5 truecm
 \item{\it Condition 5.1:} The tensor field
$\sigma_{ab}$          vanishes          throughout,         i.e.,
$\xi_{[a}R_{b]e}^{^{(4)}}\xi^e=0$.
 
\medskip \item{\it Condition 5.2:} The tensor field $\rho_{ab}$ has the
property that, in a geometrically preferred local coordinate system,
its components, $\rho_{AB}$ ($A,B=1,2$),  can be given exclusively
in terms of the induced 3-geometry, $\hat
h_{ab}$.
 
\leftskip=1 truecm
\parindent 20 pt
\medskip
 
 In  particular,  {\it  Condition  5.1}  implies  that  (at  least
locally) there exists such a function $\omega$ that $\omega_a=\hat
D_a\omega$.  Since  we are  dealing with  the case  of independent
form  fields  --  i.e.,  $(\hat  D_{[a}v)\omega_{b]}\not=0$ -- the
functions $v$ and $\omega$ are then functionally independent. {\it
Condition 5.2} might be satisfied when (3.17) can be solved for $v$,
moreover,  (by  using  the  relevant  expression  for $v$) one can
eliminate               thereby               $v$              and
${{{\psi_{_{(j)}}}^{a_1...a_m}}}_{b_1...b_n}$  from   $\rho_{AB}$.
Whenever both of  the above conditions  hold (3.19) can  be recast
into    the    form   $$H_{AB}=v^{-2}\{(\partial_Av)(\partial_Bv)+
(\partial_A\omega)(\partial_B\omega)\}, \eqno(5.1)$$ where we  used
the  expression  (4.3)  for  $H_{AB}$.   Furthermore,  due to {\it
Condition 5.2} the left hand side of (5.1) depends exclusively  on
the induced 3-metric
while the right hand side of it
depends merely on the functions  $v$ and $\omega$.  Since $v$  and
$\omega$ are functions  of $x^1$ and  $x^2$, equation (5.1)  shows
that the  same is  true for  the functions  $H_{AB}$ even  if, for
instance, some of  the components of  $\hat h_{ab}$ may  depend on
$x^3$.  (Note that this property of the functions, $H_{AB}$, is in
fact a simple  consequence of the  general result (3.13).)   Since
$v$ and  $\omega$ are  functionally independent  we have  that the
functions  $H_{AB}$  can  be  considered  as  the  components of a
non-singular Riemannian metric on a 2-dimensional manifold.   Note
that  the  right  hand  side  of  (5.1)  is  just  the  well-known
representation  of  a  Riemannian  2-metric  in  local coordinates
$(v,\omega)$  with  Gaussian  curvature  $-1$.   Hence,  for   the
Gaussian   curvature,   $K_{_{H}}$,   of   the   metric, $H_{AB}$,
$$K_{_{H}}=-1\eqno(5.2)$$ has to hold.  This equation is, in fact,
a fourth-order partial differential equation for the components of
the  tensor  fields  $\hat  h_{ab}$.   For  the  case  of linearly
independent form fields under consideration (5.2) is the necessary
and sufficient condition  for the existence  of functions $v$  and
$\omega$ satisfying (5.1).
 
The outline of the proof of the above statement can be given as follows:
 Since we are  considering the case  of linearly independent  form
fields, (3.21) and (4.4) yield
$$\partial_A\omega=\epsilon_1{H_{A2}(\partial_1v)-H_{1A}(\partial_2v)  \over
det(H_{AB})^{1\over 2}},\eqno(5.3)$$  where the ambiguity
in sign of $\omega$ is indicated by $\epsilon_1$ (i.e., $\epsilon_1=\pm1$).
Substituting (5.3) into (5.1) with setting $A,B=2$
and    solving    for    $\partial_1v$    we obtain
$$\partial_1v={H_{12}(\partial_2v)+\epsilon_2 det(H_{AB})^{1\over
2}[v^2       H_{22}-(\partial_2v)^2]^{1\over       2}        \over
H_{22}},\eqno(5.4)$$ where  $H_{22}\not=0$ since  otherwise $(\hat
D_{[a}v)\omega_{b]}$  should  vanish and $\epsilon_2=\pm 1$. Furthermore, the
substitution of
(5.4)   into   (5.3) yields   $$\partial_2\omega=\epsilon_1\epsilon_2[v^2
H_{22}-(\partial_2v)^2]^{1\over        2},\eqno(5.5)$$         and
$$\partial_1\omega=-\epsilon_1{det(H_{AB})^{1\over     2}(\partial_2v)     -
\epsilon_2H_{12}[v^2  H_{22}-  (\partial_2v)^2]^{1\over  2}  \over
H_{22}}.\eqno(5.6)$$  Equations  (5.4)  -  (5.6) are equivalent to
(5.1).    The   integrability   condition,   $\partial_2\partial_1
\omega=\partial_1\partial_2 \omega$, for the function $\omega$ can
be shown [1] to give rise to the following Appel equation 
$$2H_{22}(\partial_2\partial_2 U)-4 H_{22}(\partial_2U)^2-(\partial_2
H_{22})(\partial_2U)+H_{22}^2+\Phi [H_{22}-4(\partial_2U)^2]^{1\over 2}=0
,\eqno(5.7)$$   where $U\equiv{1\over 2}{\rm ln}(\ve v)$ and
$$\Phi\equiv{1\over 4}\epsilon_2\cdot det(H_{AB})^{-{1\over  2}}\Bigl\{-2
H_{22}(\partial_2        H_{21})+        H_{21}(\partial_2H_{22})+
H_{22}(\partial_1H_{22})\Bigr\}. \eqno(5.8)$$ Utilizing Cosgrove's
substitution (see  Ref. [1,2])  $$\partial_2U=-(H_{22})^{1\over 2}
M(1+M^2)^{-1},\eqno(5.9)$$  we  obtain  from  (5.4) and (5.7) the
following   pair    of   Riccati    equations   $$\partial_A    M=
X_A+2Y_AM+Z_AM^2.\eqno(5.10)$$ Here the  functions $X_A, Y_A$  and
$Z_A$ are  defined as  $$X_A={\epsilon_2\over 4  det(H_{AB})^{1\over
2}H_{22}}     \Bigl[     H_{12}(\partial_A    H_{22})-H_{22}\bigl(
\partial_AH_{12}+                                       \partial_2
H_{A1}-\partial_1H_{A2}\bigr)\Bigr]+{1\over                     2}
H_{A2}(H_{22})^{-{1\over    2}},    \eqno(5.11)$$    $$Y_A={
1\over
2}\epsilon_2\cdot\delta_{A1}\cdot                det(H_{AB})^{1\over
2}(H_{22})^{-{1\over             2}},\eqno(5.12)$$             and
$$Z_A=X_A-H_{A2}(H_{22})^{-{1\over    2}}.    \eqno(5.13)$$    The
integrability  conditions  for  the  simultaneous  set  of Riccati
equations, (5.10), reduce to a single condition [1,2], which,  not
unexpectedly, may be put into the form of (5.2).
 
Summarizing the results of this section we can say the following: To get a
resolution of the basic field variables we have to solve first (5.2) for
$\hat h_{\alpha\beta}$.  Then the solutions of the simultaneous Riccati
equations, (5.10), can be used to determine the function $v$ via (5.4) and
(5.9).  Afterwards, (5.3) can be applied to construct the function
$\omega$.  A detailed discussion about the resolution of the corresponding
problems for the vacuum case -- particularly, about the solutions of
Riccati equations of the above type -- can be found in Ref. [1]. Finally,
using these functions -- $v,\omega$ and $\hat h_{\a\b}$ --
the Euler-Lagrange equations have to be solved for the components of tensor
fields representing the matter content.
 
It is worth mentioning that (5.1) inherits a remarkable feature of the
corresponding equation given for the vacuum case noticed by Geroch [3].
Namely, this equation is invariant under the action of an $SL(2$,{\re}$)$ 
transformation. Two of the relevant parameters are associated with gauge 
transformations 
but there exists a one-parameter subclass of `effective' $SL(2$,{\re}$)$
transformations yielding new solutions from known ones. In particular, 
by starting with a particular solution, $(v_0,\omega_0)$,
of (5.1) associated with a fixed set of functions $H_{AB}$ one can
generate a one-parameter family of solutions, $(v_\tau,\omega_\tau)$ to 
this equation. More precisely, one can show by a straightforward 
modification of the proof of Theorem 1. of Ref. [1] that for fixed functions
$H_{AB}$ satisfying (5.2) the full set of solutions of (5.1) -- apart
from those related to gauge transformations of the spacetime,
$(M,g_{ab})$, -- is generated from the particular solution,
$(v_0,\omega_0)$, by the transformation $$v_\tau={v_0\over
(cos\tau-\omega_0 sin\tau)^2+v_0^2 sin^2\tau}, \eqno(5.14)$$
$$\omega_\tau={(sin\tau+\omega_0 cos\tau) (cos\tau-\omega_0 sin\tau)-
v_0^2 sin\tau cos\tau \over (cos\tau-\omega_0 sin\tau)^2+v_0^2
sin^2\tau}. \eqno(5.15)$$
 
 \medskip There is, however, a significant difference between  the
vacuum case  and the  case under  consideration.  Namely,  for the
case  of  vacuum  the  relevant  form  of  (5.1) is the only field
equation to be solved while  for the general case with  matter the
basic field  variables have  to satisfy,  beside (5.1), both (3.17)
and the relevant set of Euler-Lagrange  equations,
as  well.   Therefore, one would expect that
there is no matter field  so that the above transformation can be
applied.
Nevertheless,  there
exists  such  a  matter  field  (see  section  7)  where   certain
restrictions on  the basic  field variables  (associated with  the
matter content) can ensure the applicability of the transformation
(5.14)-(5.15), and, consequently, one may generate new solutions of 
Einstein's equations. In particular, this transformation  was
used to derive a number of new perfect fluid solutions from known 
ones [8,9].
 
\parindent 0 pt\bigskip
{\bf VI. Perfect fluids}
\parindent 20 pt\medskip
 
 In  this  section  some  of  the  basic  notions  and  results in
connection with perfect  fluids will be  recalled and some  of the
consequences of the presence  of Killing fields in  the spacetimes
will be discussed.
 
Consider a perfect fluid with mass density, $\rho$, and pressure,
$P$, (both quantities measured in the rest frame of the fluid),
furthermore, with 4-velocity $u^a$, where $u^au_a=-1$.  (Note that the
tensor fields ${{{\Psi_{_{(i)}}}^{a_1...a_k}}}_{b_1...b_l}$ on
$M$ for the present case are the fields $\rho, P$ and $u^a$.)  The
energy-momentum tensor is given as $$T_{ab}=\rho u_au_b +
P(g_{ab}+u_au_b),\eqno(6.1)$$ furthermore, the Euler - Lagrange equations
are $$u^a\nabla_a\rho+(\rho+P)\nabla^au_a=0,\eqno(6.2)$$
$$(\rho+P)u^a\nabla_au_b+(g_{ab}+u_au_b)\nabla^aP=0.\eqno(6.3)$$
 
 It is known  that for perfect  fluid sources these  equations are
equivalent to the `integrability' condition of Einstein's equation
$$\nabla^a T_{ab}=0.\eqno(6.4)$$  In particular,  (6.2) and  (6.3)
are equivalent to the `parallel  to $u^a$' and the `orthogonal  to
$u^a$' projections of (6.4),  respectively.  Thereby, it is  usual
in  the  formulation  of  Einstein's  theory  for  spacetimes with
perfect fluids to postulate merely the form of the energy-momentum
tensor,  $T_{ab}$,  and  solve  Einstein's  equations  since   the
equations  of  motion  for   the  fluid  then  are   automatically
satisfied.  We have chosen, however, a somewhat reversed  approach
here.  In section 2  it was assumed that  Euler-Lagrange equations
are satisfied (which implies  for the present case  that $\nabla^a
T_{ab}=0$) and this condition was used to show that some of the  basic
field equations are deducible from the others.  It is important to
emphasize that  we earn more than  we loose  by replacing  the two
basic  field  equations,  (2.11)  and  (2.12),  by  Euler-Lagrange
equations.  Equations (2.11) and  (2.12) are second order  partial
differential equations  while the above Euler-Lagrange  equations
are first order ones for perfect fluid.
 
Consider now the consequences of the presence  of
a Killing field, $\xi^a$, for
perfect fluid matter sources. First of all, $${\cal L}_\xi T_{ab}=0.
\eqno(6.5)$$
Again, by the presence of a preferred vector field, $u^a$, one might
consider the unique decomposition of ${\cal L}_\xi T_{ab}$ into symmetric
tensor fields so that each of these tensor fields has definite
`tangential' or `perpendicular' character with regard to their free
indices. Since ${\cal L}_\xi T_{ab}$ vanishes all of these projections
must vanish, as well. Thereby $({\cal L}_\xi T_{ab})u^a u^b=0$ which gives
that $${\cal L}_\xi \rho=0.\eqno(6.6)$$ Then $({\cal L}_\xi T_{ab})u^a
{\pi^b}_e=0$ yields that $${\cal L}_\xi u^a=0,\eqno(6.7)$$ or $\rho+P=0$.
(Note, however, when the  equation of  state is  chosen to  be $\rho+P=0$ 
then the energy-momentum tensor is of the form $T_{ab}=Pg_{ab}$, and (6.4)
implies that $P$  is constant throughout.   This is precisely  the
case of vacuum  fields with non-zero  cosmological constant so it seems to
be reasonable to assume  that  $\rho+P$  is  not identically zero, and 
hereafter we do that.)  Finally, from $({\cal
L}_\xi T_{ab}){\pi^a}_e{\pi^b}_f=0$ we get $${\cal L}_\xi P=0\eqno(6.8)$$
throughout, where the projector, ${\pi^a}_b$, is defined to be
${\pi^a}_b={\delta^a}_b-u^au_b$. All in all, each of the physical
quantities related to the perfect fluid are invariant under the action of
the isometry group associated with $\xi^a$. Consequently, for a general
perfect fluid spacetime possessing a non-null Killing field we can use,
without loss of generality, instead of the fields $\rho, P, \ u^a$
(``${{\Psi_{_{(i)}}}^{a_1...a_k}}_{b_1...b_l}$")  given on $M$ the fields
$\rho, P,\ u_\|=
u^a\xi_a, \ u^a_\bot={h^a}_bu^b$
(``${{\psi_{_{(j)}}}^{a_1...a_m}}_{b_1...b_n}$") defined on $\cal S$.
 
Determine now the relevant form of $\rho_{ab}$ and $\sigma_{ab}$.
According      to      (2.8)      and      (6.1)      we      have
$$R^{^{(4)}}_{ab}=8\pi\bigl[(\rho+P)     u_au_b     +{1\over    2}
(\rho-P)g_{ab}\bigr],\eqno(6.9)$$ furthermore,  by the  definition
of            $\rho_{ab}$            and             $\sigma_{ab}$
$$\rho_{ab}=8\pi\Bigl[(\rho+P)\bigl\{  ({h_a}^m  u_m)({h_b}^n u_n)
+\ve          v^{-2}(u^e\xi_e)^2{\hat           h_{ab}}\bigr\}+\ve
v^{-1}(\rho-P){\hat         h}_{ab}\Bigr],\eqno(6.10)$$         and
$$\sigma_{ab}=-8\pi\epsilon_{abmn}\xi^mu^n(\rho+P)
(u^e\xi_e),\eqno(6.11)$$ hold.
 
 For simplicity, one may restrict  ones considerations  to the  case of
vanishing    $\sigma_{ab}$.     Equation    (6.11)    implies  that
$\sigma_{ab}=0$   whenever   either   of   the   following  hold:
$u_e\xi^e=0$, or $\xi^{[a}u^{b]}=0$ (or $\rho+P=0$ but this case has been 
excluded earlier).  
Thereby, we can say that $\sigma_{ab}=0$ throughout if and only if either
the 4-velocity of the fluid, $u^a$, is parallel to the Killing  field,
$\xi^a$,  which  means  that  the  spacetime  is  stationary   and
$$u^a=(-v)^{-{1\over       2}}\xi^a,\eqno(6.12)$$            or
$$u^a\xi_a=0,\eqno(6.13)$$  which  might  be  the case whenever the
Killing field,  $\xi^a$, is  spacelike.  For  both of  these cases
{\it  Condition  5.1}  holds  which  implies that there exists (at
least locally) a function  $\omega$ such that $$\omega_a=\hat  D_a
\omega.\eqno(6.14)$$
 
Let us consider  the following particular  case of perfect  fluid
sources: There are two commuting Killing fields,  $\xi^a_{_{(A)}}$
($A=1,2$),  on  the  spacetime  and  the  4-velocity of the fluid,
$u^a$,  can  be  given  as  a  linear combination of these Killing
fields      $$u^a={\cal      A}\bigl(\xi^a_{_{(1)}}+{\cal      B}\
\xi^a_{_{(2)}}\bigr).\eqno(6.15)$$ Then with linearly  independent
Killing fields (6.7) and (6.15) yield that the functions $\cal A$
and  $\cal  B$  satisfy  $${\cal  L}_{\xi_{_{(A)}}}{\cal  A}={\cal
L}_{\xi_{_{(A)}}}{\cal  B}  =0\  \  \  (A=1,2).\eqno(6.16)$$  Now,
applying (6.7) and (6.8) for the Killing fields,
$\xi^a_{_{(A)}}$,   and   using   (6.15)   we   get   that $${\cal
L}_u\rho={\cal  L}_u  P=0,\eqno(6.17)$$  and,  the  equations   of
motion,       (6.2)        and       (6.3),        reduce       to
$$(\rho+P)\nabla^au_a=0,\eqno(6.18)$$
$$(\rho+P)u^a\nabla_au_b+\nabla_bP=0.\eqno(6.19)$$
 
 One extracts  from (6.15)  - (6.16)  that the  fluid is expansion
free, i.e.,  $\nabla_a u^a=0$  throughout.  Thereby,  (6.18) holds
identically.   Furthermore,  a  straightforward calculation yields
that       $$u^a\nabla_a        u_b=       -{1\over        2}{\cal
A}^{2}\Bigl\{\nabla_b(-{\cal    A}^{-2})    -{\partial     (-{\cal
A}^{-2})\over    \partial     {\cal    B}}\nabla_b     {\cal    B}
\Bigr\}.\eqno(6.20)$$  which  along  with  (6.19) and (6.20) gives
that $$\nabla_a P +{1\over 2}(\rho+P)\Bigl[\nabla_a({\rm  ln}{\cal
A}^{-2}) -{\partial  ({\rm ln}{\cal  A}^{-2})\over \partial  {\cal
B}}\nabla_a {\cal  B} \Bigr]=0.\eqno(6.21)$$  As it  was argued in
Ref.   [10],   (6.21)   implies   that   $P=P({\cal   A,B})$    and
$\rho=\rho({\cal  A,B})$  even  if  $\cal  A$  and  $\cal  B$  are
functionally  dependent  or  constant.   Furthermore,  since   the
4-velocity --  given by  (6.15) --  is a  unit timelike vector we
have $${\cal  A}^{-2}=-\bigl\{(\xi^a_{_{(1)}}\xi_{_{(1)}a})+2{\cal
B}                             (\xi^a_{_{(1)}}\xi_{_{(2)}a})+{\cal
B}^2(\xi^a_{_{(2)}}\xi_{_{(2)}a})\bigr\},\eqno(6.22)$$ which along
with (6.21) (and the above conclusion) gives that the equation  of
state  must  be  of  the  form  $$\rho=\rho(P).\eqno(6.23)$$   The
remained  Euler  -  Lagrange  equation, (6.21), simplifies further
whenever ${\partial  ({\rm ln}{\cal  A}^{-2})\over \partial  {\cal
B}}\nabla_a {\cal B} =0,$ i.e., $$\nabla_a {\cal B} =0 \ \ \ \ \ \
{\rm  or}\  \  \  \  \  \ \ {\partial ({\rm ln}{\cal A}^{-2})\over
\partial {\cal B}}=0.\eqno(6.24)$$  The case $\nabla_a{\cal  B}=0$
is  that  of  a  `rigid  fluid',  i.e.,  the 4-velocity, $u^a$, is
parallel to the timelike Killing field $\xi^a=\xi^a_{_{(1)}}+{\cal
B} \xi^a_{_{(2)}}$.  It is  important to emphasize that  equations
(6.15) - (6.23)  along with their  consequences hold (with  ${\cal
B}=0$) without any alteration even if the spacetime admits only  a
single timelike Killing field, $\xi^a=\xi^a_{_{(1)}}$, parallel to
the 4-velocity of the fluid, $u^a$.
 
 The other  possibility, ${\partial  ({\rm ln}{\cal  A}^{-2})\over
\partial {\cal B}}=0$, along with (6.22) gives that
${\cal B}=-{
\xi^e_{_{(1)}}\xi_{_{(2)}e}\over  \xi^f_{_{(2)}}\xi_{_{(2)}f}  }$,
i.e.,  the  4-velocity  of  the  fluid,  $u^a$,  is  orthogonal to
$\xi^a_{_{(2)}}$  which,  therefore,  must  be a spacelike Killing
field.
 
 Note that for both of  these cases not merely the  Euler-Lagrange
equations are simplified  but, in accordance  with this fact,  the
potential space associated with the Lagrangian of this  particular
case of `gravity plus perfect fluid' system admits a symmetry [10].
 
Moreover, equations (6.21), (6.23) and (6.24) yield then that $${\cal
A}^{-2}(P)={\cal A}^{-2}_0\cdot exp\Bigl[-2\int^P_{P_0} {dP'\over
{\rho(P')+P'}}\Bigr],\eqno(6.25)$$ where ${\cal A}^{-2}_0$ and $P_0$ are
constants of the integration. Consequently, whenever the 4-velocity of the
fluid, $u^a$, is either parallel to a Killing field, $\xi^a$, or spanned
by two commuting Killing fields, as in (6.15), with ${\cal B}=-{
\xi^e_{_{(1)}}\xi_{_{(2)}e}\over \xi^f_{_{(2)}}\xi_{_{(2)}f} }$, and, the
equation of state, $\rho=\rho(P)$, is known then the function ${\cal
A}^{-2}={\cal A}^{-2}(P)$ or $P=P({\cal A}^{-2})$ can be determined via
(6.25).  Note that the function ${\cal A}^{-2}$ possesses the form
$$\eqalign{ {\cal A}^{-2} =\cases{ -v,& if $ u^{[a}\xi^{b]}=0$;\cr W^2
v^{-1},& if $u^a\xi_{_{(2)}a}=0,$\cr} }\eqno(6.26)$$ where
$$W^2={-(\xi^e_{_{(1)}}\xi_{_{(1)}e})(\xi^f_{_{(2)}}\xi_{_{(2)}f})+
(\xi^h_{_{(1)}}\xi_{_{(2)}h})^2}.\eqno(6.27)$$ Note that the function $W$
has the following simple geometrical meaning. In canonical Weyl
coordinates, $(\rho,z,\phi)$, the 3-metric, $\hat h_{ab}$, can be given as
$$\hat h_{\a\b}=diag\{Exp(2\g),Exp(2\g),-W^2\},\eqno(6.28)$$ where $\g$
and $W$ are functions of the coordinates $(\rho,z)$ [4].
 
\parindent 0 pt\bigskip
{\bf VII. Perfect fluids with 4-velocity parallel to a Killing
field}
\parindent 20 pt\medskip

 In this section  we shall apply  the results of  the previous
sections for perfect fluid spacetimes possessing a timelike
Killing field, $\xi^a$, parallel  to the 4-velocity of  the fluid,
$u^a$.  Such a fluid has expansion- and shear-free flow, i.e.,  it
is `rigid'. Thereby one might ask whether there exists any physically
realistic situation in which such a model can be applied.
However, it was shown by Geroch and Lindblom [11] that in a generic theory
of relativistic dissipative fluids the equilibrium states
are perfect fluid states. Furthermore, they showed that for these perfect
fluids -- which represent the equilibrium configurations of dissipative
relativistic fluids --
the  4-velocity is parallel
to a Killing field [11]. Therefore, the model we are dealing with in this
section has to have physical relevance, and, in fact, it is the needed one
as long as we are looking for a faithful description of possible equilibrium
configurations of relativistic dissipative fluids.
 
First the applicability of the generalization of Cosgrove's method
then the general approach will be considered.
Clearly, for this type of perfect fluids, {\it Condition 5.1} is
satisfied and
we show that {\it Condition 5.2} holds, as well.  Now, since
$u^{[a}\xi^{b]}=0$, the Killing field, $\xi^a$, is timelike so $\ve$
takes the value $-1$. Furthermore, (6.6) yields that
$$\rho_{ab}=16\pi v^{-1}P{\hat h}_{ab}.\eqno(7.1)$$ The
relevant form of (3.17) $$\hat R_{3\b}= 16 \pi v^{-1} P \hat
h_{3\b}, \eqno(7.2)$$ can then be solved for $v$.
 Since we have a non-vanishing spacelike
vector field, $k^a$, and the 3-metric , $\hat h_{\a\b}$, is non-singular,
$\hat h_{33}$ cannot vanish.  Whenever there is another
non-vanishing one among the functions, $\hat h_{3\b}$, then (7.2) gives rise
to an algebraical restriction on the components $\hat R_{3\beta}$ of the
Ricci tensor associated with $\hat h_{ab}$.   We
obtain from (7.1) and (7.2) $$\rho_{AB}= {\hat R_{33}\over \hat h_{33}}\hat
h_{AB},\eqno(7.3)$$ which means that {\it Condition 5.2} is
satisfied. Furthermore, this equation yields, along with (4.3),
$$H_{AB}=2(\hat
R_{AB}-{\hat R_{33}\over \hat h_{33}} \hat h_{AB}) .\eqno(7.4)$$
 
For the particular case under consideration the functions $H_{AB}$ depend
exclusively on the induced 3-metric, $\hat h_{ab}$, and the relevant form
of (5.2) is, in fact, a fourth order partial differential equation for the
components of $\hat h_{ab}$.  It is striking to what an extent the
corresponding basic field equations are similar in structure to the vacuum
counterparts.  Turning back to the main issue, note that the functions $v$
and $\omega$ can be determined by virtue of (5.3)-(5.13).  Finally, the
pressure, $P$, can be determined by (7.2) and the mass density, $\rho$, by
the Euler-Lagrange equations, (6.21).
 
After solving (5.2) and fixing the functions $H_{AB}$ we may ask for the
conditions under which the transformation (5.14)-(5.15) yields new solutions of
the basic field equations. The two equations to be solved are, for the
present case, (7.2) and (6.21).  It is straightforward to check that by
choosing $P_\tau$ to satisfy the equation $P_\tau v_\tau^{-1}=P_0v_0^{-1}$ and
deriving $\rho_\tau$ -- for each pair of the functions
$v_\tau$ and $P_\tau$ -- in virtue of (6.21) we get a one-parameter family of
solutions of the basic field equations. Note that the invariance properties of the
Lagrangian of electrically charged rigid perfect fluids was studied earlier 
by Kramer, Neugebauer and Stephani [4,12]. However, their considerations were restricted 
to the case of a static spacetime exclusively.  
Thereby the transformation transparented by eqs. (5.14) - (5.15) is new for perfect fluids
although it is a straightforward generalization of a known algorithm. 
Further analyzes related to this transformation and derivation of new perfect solutions 
by making use of it can be found in Ref. [8,9].  

An interesting subcase of these perfect fluid spacetimes, discovered
many years ago by Ehlers [12,4], is the case of vanishing pressure,
i.e., of a stationary spacetime with dust possessing 4-velocity, $u^a$,
parallel to a timelike Killing field, $\xi^a$.  According to
(3.17), (3.18) and (7.1) the field equations are then the same as
for the vacuum case.  Note, however, that (6.21) implies then
$v=const$.  The value of $v$ may be chosen  throughout to be -1, so the
field equations simplify to $$\hat R_{3\beta}=0, \eqno(7.5)$$ $$\hat
R_{AB}={1\over 2} (\partial_A \omega)(\partial_B \omega). \eqno(7.6)$$
For the energy density, $\rho$, we  obtain
from $(2.\a.2)$, (6.9) and (6.12)  the constraint $$\rho={1\over 8 \pi}
(\partial^A
\omega)(\partial_A \omega).\eqno(7.7)$$ Accordingly, a stationary dust
({\it  sd}) solution --
represented by $v^{^{(sd)}}=-1$, $\omega^{^{(sd)}}={\rm ln}[
v^{^{(sv)}}]$ and $\hat h^{^{(sd)}}_{ab}=\hat{h}^{^{(sv)}}_{ab}$ -- can
be assigned  to every static
vacuum ({\it sv}) solution -- given in terms of $v^{^{(sv)}}$ and
$\hat{h}^{^{(sv)}}_{ab}$ -- where the energy density of the dust satisfies the
constraint (7.7) [13,4].
 
 Although the above general method can be applied in a straightforward way
to get solutions of  Einstein's equations for the selected perfect fluid
source there is  an unfavorable aspect  of this method.   Namely,
the  most  significant  physical  quantities  -- the mass density,
$\rho$, and the pressure, $P$, characterizing the possible physical states of
the fluid -- can be determined  only at the very end of  the entire
process in  terms of  the function  $v$ and  the 3-geometry, $\hat
h_{ab}$. Therefore the equation  of  state,  $\rho=\rho(P)$, has to be in
accordance of the corresponding constraints, which implies that it cannot be
chosen freely.  The general approach, introduced in section 4, can  be
used to  cure this  problem but  the price  we have  to pay is the
appearance of extra non-linearities.
 
The general method was developed to ensure
more control on the physical  properties
of matter  fields in solving the relevant set of field equations.
The  significance of  the seemingly  technical
differences between the two methods can be transparented for the
examined perfect fluid sources
as follows: Remember that the basic point in the general approach
was the specification of the functional form of $v$ in terms of the
3-geometry and tensor fields representing the matter content. Moreover,
for the present situation the basic field equations are (5.1) -- where
$H_{AB}$ is given by (7.4) --, (7.2)
and (6.25). Note that (6.25) -- which is, indeed, the integrated form of the
Euler-Lagrange equation --
gives now the functional relation between the functions $v,\rho$ and
$P$ as
$$v(P)=v_0 \cdot exp\Bigl[-2\int^P_{P_0} {dP'\over
{\rho(P')+P'}}\Bigr].\eqno(7.8)$$ With the help of (7.2)
and (7.8) it can be shown that the function $v$ can be given in terms 
of the 3-geometry
exclusively. Note that (7.8) is independent of the 3-geometry and, indeed,
it is  the
only equation where one can control the physical properties of the
solution describing the fluid
by the substitution of a physically realistic equation of state. Consider, for
instance, the case of  polytrope equation of
state, i.e., $$\rho=cP^\a,\eqno(7.9)$$ where $\a\in(0,1)\cup(1,\infty)$ 
and $c>0$.
Using then (7.8) and
(7.9) one gets $$v=v_0\biggl[{{P^{\a-1}+c}\over{P^{\a-1}_0+c}}\biggr]^{-{2
\over {\a-1}}}.\eqno(7.10)$$
After solving (7.2) for $v$ and substituting the resulted
function into (5.1)
one might attempt to get solutions of the yielded equation,
where now $H_{AB}$ is given by (7.4). Remember 
that for the present case of perfect
fluids $\sigma_{ab}$ is identically zero thereby one has to solve merely the
relevant form of (4.5) and (4.6) along with the possible two equations
involved by (7.2) [Note that the last two equations might give further
algebraical relationships
between certain
components
of the 3-metric and the 3-Ricci tensor.]
Although the derivation of the equations
is straightforward the appearance of extra non-linearities -- related to
the functional form of $v$ -- are
frightening. Note, however, that whenever one is able to find solutions of
these
equations the physical relevance of the solutions is automatically assured.
 
\parindent 0 pt\bigskip
{\bf VIII. Final remarks}
\parindent 20 pt\medskip
 
A new formulation of Einstein's equations for spacetimes admitting a
non-null Killing vector field and arbitrary matter field was
given in this paper. First it was shown that some of the basic field equations
are always deducible from the others. Then the existence of a geometrically
preferred vector field and related coordinate systems were shown. Based on
the associated simplifications, two methods were presented obtaining systems of
partial differential equations for the basic variables associated with the
spacetime geometry and with the matter content.
Both of the developed approaches
were applied for perfect fluid spacetimes which describe equilibrium
configurations of relativistic dissipative fluids. The symmetry properties
of the relevant equations and differences of the two approaches were studied.
It was shown, furthermore, that the techniques which were developed
by Cosgrove [1] for the vacuum stationary axisymmetric problem can be
generalized straightforwardly
for these perfect fluid spacetimes despite the fact that in our examinations
we assumed  merely the existence of a single timelike Killing field.
 
It is worth emphasizing that the general results of this paper, given in
details in sections 2 -- 5, are valid for any spacetime in Einstein's
theory which possesses a non-null Killing field  and essentially arbitrary
matter fields. Thereby, it would deserve further studies to find out how
to apply these results for even more interesting situations in which time
dependence may occur and/or different types of matter fields are
present.

%\vfill
%\eject
 
\bigskip
{\bf Acknowledgments}
\medskip
 
This research was
supported in parts by the OTKA grants F14196 and T016246. I would like
to say thank to Dr. Tam\'as Dolinszky for careful reading of the manuscript and
the Ervin Schr\"odinger Institute in Vienna for its hospitality during work 
on this paper.
 
\bigskip
{\bf References}
 
\item{[1]} C.M. Cosgrove: J. Phys.  A {\bf 11}, 2389 (1978)
 
\item{[2]} E.D. Fackerell and R.P. Kerr: Gen. Rel. Grav. {\bf
23}, 861 (1991)
 
 \item{[3]} R. Geroch: J. Math.  Phys. {\bf 12}, 918 (1971)
 
\item{[4]} D. Kramer, H. Stephani, M. McCallum and E. Herlt: {\it Exact
solutions of Einstein's equations}  (Cambridge; Cambridge
University Press, 1980)
 
 \item{[5]} S. Harris: Class.  Quant.  Grav. {\bf 9}, 1823 (1992)

\item{[6]} B. Luk\'acs and Z. Perj\'es: in {\it Proc. 1st. Marcel
Grossmann Meeting}, Ed. R. Ruffini, (North-Holland 1976) p. 281

\item{[7]} C. Hoenselaers: Prog. Theor. Phys. {\bf 57}, 1223
(1977)
 
\item{[8]} I. R\'acz and J. Zsigrai: Class. Quant. Grav. {\bf 13}, 2783
(1996)

\item{[9]} I. R\'acz and J. Zsigrai: Class. Quant. Grav. {\bf 14} 1997 (1997)

\item{[10]} H. Stephani and R. Grosso: Class. Quant. Grav. {\bf
6}, 1673 (1989)
 
\item{[11]} R. Geroch and L. Limdblom: Ann. Phys. {\bf 207}, 394
(1991)
 
\item{[12]} D. Kramer, G. Neugebauer and H. Stephani:
 Fortschr, Physik. {\bf 20}, 1 (1972)

\item{[13]} J. Ehlers: in {\it Colloques
Internationaux C.N.R.S. No. 91} (Les th\'eoties relativistes de la
gravitation, {\bf 275}, 1962)

\vfill\eject
\end